\documentclass[journal = ancac3]{achemso}
\setkeys{acs}{usetitle = true}

\usepackage{graphicx}
\usepackage{color}
\usepackage{amsfonts}
\let\oldAA\AA
\renewcommand{\AA}{\text{\normalfont\oldAA}}

\title{Ultrafast Band Structure Control of a Two-Dimensional Heterostructure}
\author{S{\o}ren~Ulstrup}
\affiliation{Advanced Light Source, E. O. Lawrence Berkeley National Laboratory, Berkeley,
California 94720, USA}
\author{Antonija Grubi\v{s}i\'{c} \v{C}abo}
\author{Jill A. Miwa}
\affiliation{Department of Physics and Astronomy, Interdisciplinary Nanoscience Center, Aarhus University,
8000 Aarhus C, Denmark}
\author{Jonathon M. Riley}
\affiliation{SUPA, School of Physics and Astronomy, University of St. Andrews,
St. Andrews, United Kingdom}
\author{Signe S. Gr{\o}nborg}
\affiliation{Department of Physics and Astronomy, Interdisciplinary Nanoscience Center, Aarhus University,
8000 Aarhus C, Denmark}
\author{Jens C. Johannsen}
\affiliation{Institute of Condensed Matter Physics, \'Ecole Polytechnique F\'ed\'erale de Lausanne (EPFL), Switzerland}
\author{Cephise Cacho}
\author{Oliver Alexander}
\author{Richard T. Chapman}
\author{Emma Springate}
\affiliation{Central Laser Facility, STFC Rutherford Appleton Laboratory, Harwell, United Kingdom}
\author{Marco Bianchi}
\author{Maciej Dendzik}
\affiliation{Department of Physics and Astronomy, Interdisciplinary Nanoscience Center, Aarhus University,
8000 Aarhus C, Denmark}
\author{Jeppe V. Lauritsen}
\affiliation{Department of Physics and Astronomy, Interdisciplinary Nanoscience Center, Aarhus University,
8000 Aarhus C, Denmark}
\author{Phil D. C. King}
\affiliation{SUPA, School of Physics and Astronomy, University of St. Andrews,
St. Andrews, United Kingdom}
\author{Philip~Hofmann}
\affiliation{Department of Physics and Astronomy, Interdisciplinary Nanoscience Center, Aarhus University,
8000 Aarhus C, Denmark}
\email{philip@phys.au.dk}

\begin{document}

\newpage

\begin{abstract}
The electronic structure of two-dimensional (2D) semiconductors can be significantly altered by screening effects, either from free charge carriers in the material itself, or by environmental screening from the surrounding medium. The physical properties of 2D semiconductors placed in a heterostructure with other 2D materials are therefore governed by a complex interplay of both intra- and inter-layer interactions. Here, using time- and angle-resolved photoemission, we are able to isolate both the layer-resolved band structure and, more importantly, the transient band structure evolution of a model 2D heterostructure formed of a single layer of MoS$_2$ on graphene. Our results reveal a pronounced renormalization of the quasiparticle gap of the MoS$_2$ layer. Following optical excitation, the band gap is reduced by up to $\sim\!$400 meV on femtosecond timescales due to a persistence of strong electronic interactions despite the environmental screening by the $n$-doped graphene. This points to a large degree of tuneability of both the electronic structure and electron dynamics for 2D semiconductors embedded in a van der Waals-bonded heterostructure.
\end{abstract}

\newpage

\maketitle
Van der Waals-bonded heterostructures of two-dimensional (2D) atomic sheets hold great promise for the bottom-up design of materials with new properties \cite{Geim:2013aa,Georgiou:2013aa,Roy:2013aa}. Realising heterostructures with desired functionality, however, remains a formidable challenge. Environmental screening from neighbouring layers can severely modify the band structures of the individual 2D materials, even though no real chemical bonds are formed between them \cite{Ugeda2014}. In 2D semiconductors the quasiparticle band gap and the exciton binding energy have been observed to be strongly influenced both by the choice of substrate material and by excited electrons within the 2D material \cite{Steinhoff:2014aa,Chernikov:2015aa,Chernikov:2015ab,Pogna:0aa}. When placing a single layer transition metal dichalcogenide (SL-TMDC) on a metallic substrate such as Au(111), a strong band gap renormalization is observed but the reduced band gap is almost unaffected by an additional excited carrier density \cite{Antonija-Grubisic-Cabo:2015aa,Bruix:2016}. When the SL-TMDC is placed on a weakly screening insulator such as SiO$_2$, the static band gap is close to that expected for the free-standing case \cite{makatomically2010,splendiani2010,Qiu:2013} but optically excited carriers can give rise to a giant band gap reduction \cite{Chernikov:2015aa}. 
Such carrier-induced renormalization effects are somewhat similar to what is observed in bulk semiconductors  \cite{Kalt:1992aa} and quantum wells \cite{Trankle:1987aa,Das-Sarma:1990aa}, where they can be used to generate materials with non-linear optical properties. A key advantage of using a truly 2D semiconductor, however, is that its environment can be freely chosen, which should greatly enhance the tuneability of the system. We demonstrate this capability for a heterostructure of SL MoS$_2$ and graphene (MoS$_2$/G). Screening of the semiconductor by the adjacent graphene is expected to induce a moderate reduction of the quasiparticle band gap \cite{Ugeda2014,Jin:2015}, yet we find that the MoS$_2$ layer retains a strong optical tuneability. Using time- and angle-resolved photoemission (TR-ARPES), we not only directly visualise how this causes the band structure to change on femtosecond time scales following optical excitation of free charge carriers, but are also able to separate the underlying electronic structure and carrier dynamics from the SL MoS$_2$ and the graphene.

In TR-ARPES, electrons are excited into unoccupied states using low-energy photons and then photoemitted with a time-delayed higher energy laser pulse. This type of experimental set-up provides information on the band structure and carrier population in the equilibrium and excited states, and on the time dependent population change after the initial excitation. For the material system and moderate excitation studied here, one does not typically expect the pump pulse to create changes in the band structure as such, but merely in the population of the states. This is indeed what we find for the graphene layer in our MoS$_2$/G heterostructure: Electrons are pumped into the conduction band (CB) and follow a decay dynamics that closely resembles earlier findings for epitaxial graphene \cite{Johannsen:2015}. On the other hand, the adjacent MoS$_2$ bands are rigidly shifted in the presence of the excited carriers. We determine how these band shifts lead to a significant reduction of the quasiparticle band gap as a function of the excited carrier density, consistent with a recent theoretical calculation \cite{Steinhoff:2014aa}.

\section{Results and Discussion}

\begin{figure*} [ht!]
\begin{center}
\includegraphics[width=1\textwidth]{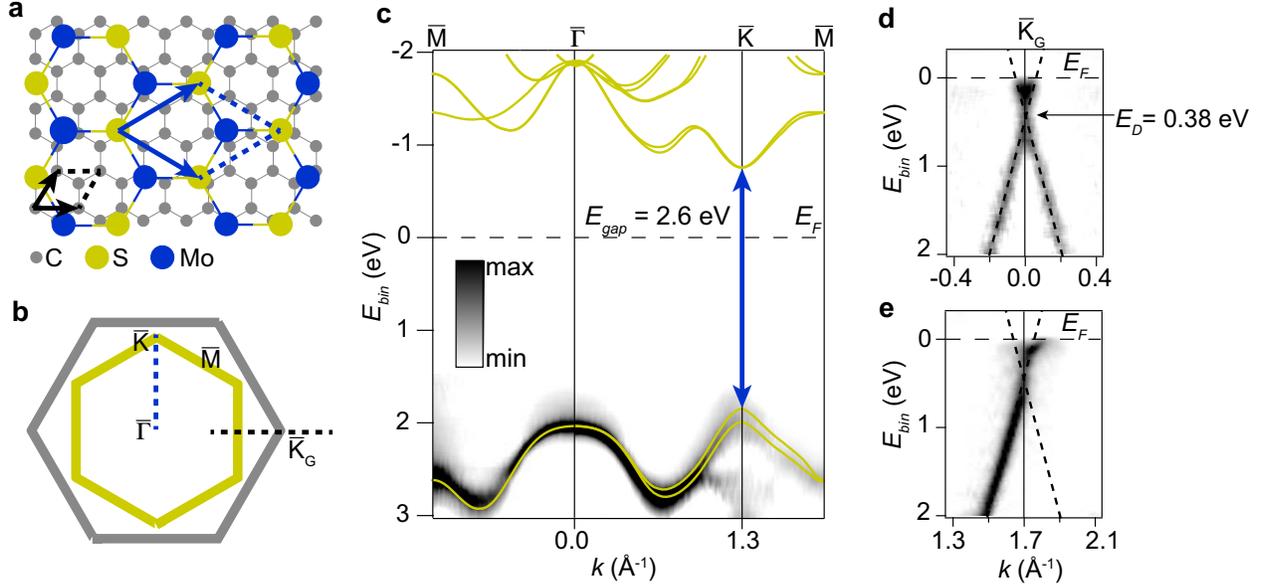}
\caption{Equilibrium electronic structure of SL MoS$_2$ on graphene: (a) Alignment of the crystal lattices of epitaxial SL MoS$_2$ on graphene. The unit cells (arrows and dashed lines) are rotated by 30$^{\circ}$ with respect to each other. (b) Corresponding BZs for MoS$_2$ (yellow) and graphene (grey). The measurement directions for the TR-ARPES experiments are shown by blue (SL MoS$_2$)  and black (graphene) dashed lines. The corner of the graphene BZ is labeled $\bar{K}_G$, while the MoS$_2$-related symmetry points do not have an index. (c) Band structure of the upper VB for SL MoS$_2$ measured by ARPES and overlaid with the theoretical VB and CB (yellow curves) assuming a quasiparticle gap of 2.6~eV for SL MoS$_2$/G. (d)-(e) Band structure around the Dirac point ($E_D$) of graphene with a linear dispersion (black dashed lines) overlaid as a guide to the eye. The cuts in (d) and (e) are perpendicular and parallel to the dashed line in (b), respectively. The cut in (e) corresponds to the measurement direction in TR-ARPES. The second derivative of the photoemission intensity is shown for all the ARPES data.}
\label{fig:1}
\end{center}
\end{figure*}

The MoS$_2$/G heterostructure was grown on a silicon carbide substrate by van der Waals epitaxy (see Materials and Methods section). Due to the weak interaction between the individual 2D layers the MoS$_2$ grows with many rotational domains. However, our synthesis method yields a preference for MoS$_2$ domains that can be rotated by 30$^{\circ}$ and 90$^{\circ}$ with respect to the underlying graphene (see sketch of the 30$^{\circ}$ case in Figure \ref{fig:1}a). As described in detail in Ref. \citenum{miwavander2015} we can determine this rotational preference directly from ARPES scans of the overlapping BZs of the two materials since both the Dirac cone of graphene and the valence band maximum (VBM) of SL MoS$_2$ are found at the $\bar{K}$ point of their respective BZs (see Figure \ref{fig:1}b). Note that since the electronic bands of MoS$_2$ domains rotated by 30$^{\circ}$ and 90$^{\circ}$ overlap we can not distinguish such domains in ARPES. The rotation of the graphene and MoS$_2$ BZs allows us to separately access the excited carrier dynamics around the Dirac cone in graphene and around the direct band gap of SL MoS$_2$. Figure \ref{fig:1}c provides the MoS$_2$ valence band (VB) dispersion measured by ARPES with theoretical band structure calculations for the free-standing case overlaid. The Dirac cone of the underlying graphene is seen in Figure \ref{fig:1}d. The Dirac point is found at $E_D=0.38$~eV, corresponding to a strong $n$-doping ($n\approx 1.1\times10^{13}$~cm$^{-2}$) that is similar to previously reported band structure measurements for graphene synthesised by similar methodologies but without the SL MoS$_2$ film on top \cite{Bostwick:2007aa,Johannsen:2015}.

\begin{figure*} [t!]
\begin{center}
\includegraphics[width=0.8\textwidth]{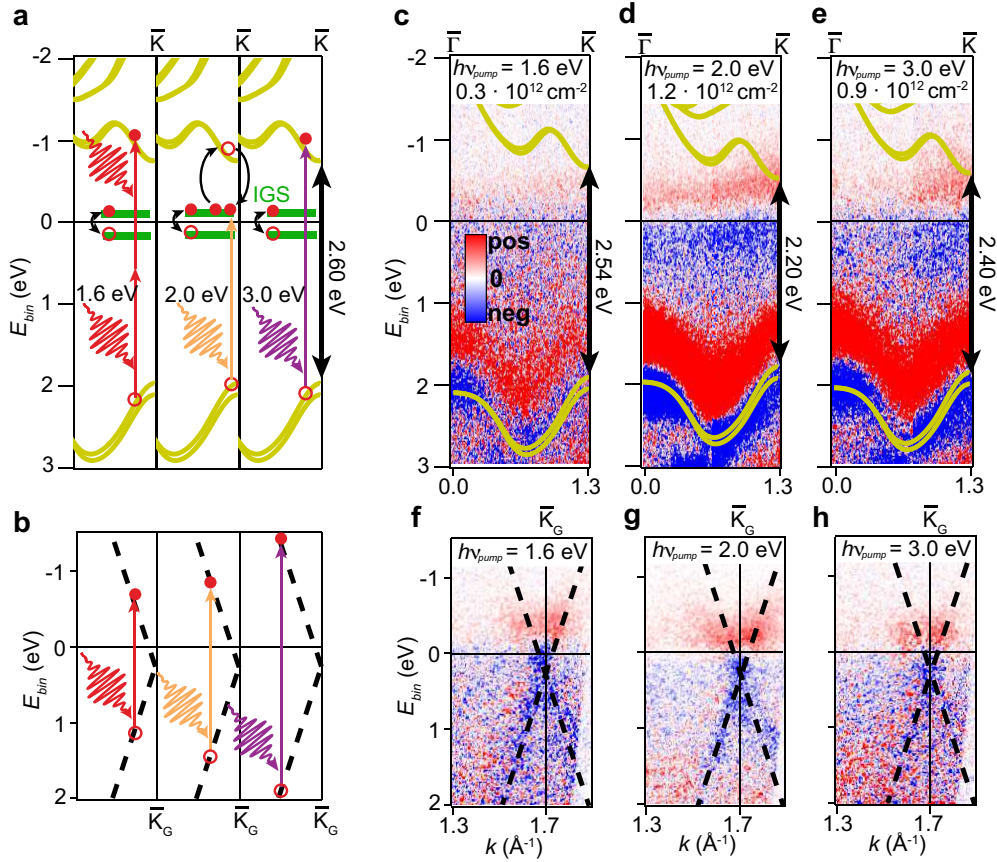}
\caption{Excitation of a MoS$_2$/G heterostructure: (a) Possible electron-hole pair excitation processes in SL MoS$_2$ for the applied pump energies of 1.6~eV, 2.0~eV and 3.0~eV. When the pump energy is below the band gap, the excitation must happen \textit{via} higher order multiple photon processes (shown for the 1.6~eV case), the involvement of in-gap states (IGS) (shown for the 2.0~eV case) or exciton dissociation (not shown). The equilibrium quasiparticle gap energy is assumed to be 2.60~eV. (b) Corresponding excitations in graphene. Direct transitions are allowed for all photon energies considered. (c)-(e) Intensity difference spectra of the MoS$_2$ bands at peak excitation along the $\bar{\Gamma}-\bar{K}$ direction. Values for the renormalized gap are provided based on the VB shift and assuming a symmetric CB shift. The maximum optically induced hole density is provided in each panel.  (f)-(h)  Intensity difference in the graphene bands at  peak excitation. The fluences are (c), (f) $F=0.7$~mJ/cm$^2$, (d), (g) $F=3.0$~mJ/cm$^2$ and (e), (h) $F=1.3$~mJ/cm$^2$.}
\label{fig:2}
\end{center}
\end{figure*}

The unoccupied electronic states and band gap for SL MoS$_2$ are not accessible by ARPES. For the sketches in Figures \ref{fig:1}c and \ref{fig:2}a, we assume a static band gap of 2.6~eV. This value is estimated from the theoretical values of 2.7-2.8~eV for a free-standing layer \cite{Huser:2013aa,Qiu:2013,Steinhoff:2014aa} and an environmental screening-induced renormalization of $\approx 0.2$~eV that has been observed for the similar system of MoSe$_2$ on bilayer graphene \cite{Ugeda2014}. We shall later see that a static band gap of this size fits well with our TR-ARPES measurement of the excited state when we extrapolate our measured band gap renormalization to the limit of zero free carrier density. Note that the static band gap renormalization for SL MoS$_2$ on graphene is substantially smaller than for MoS$_2$ on a truly metallic substrate such as Au(111) where the band gap is reduced by $\approx 0.9$~eV \cite{Antonija-Grubisic-Cabo:2015aa,Bruix:2016}.

\begin{figure*} [t!]
\begin{center}
\includegraphics[width=0.8\textwidth]{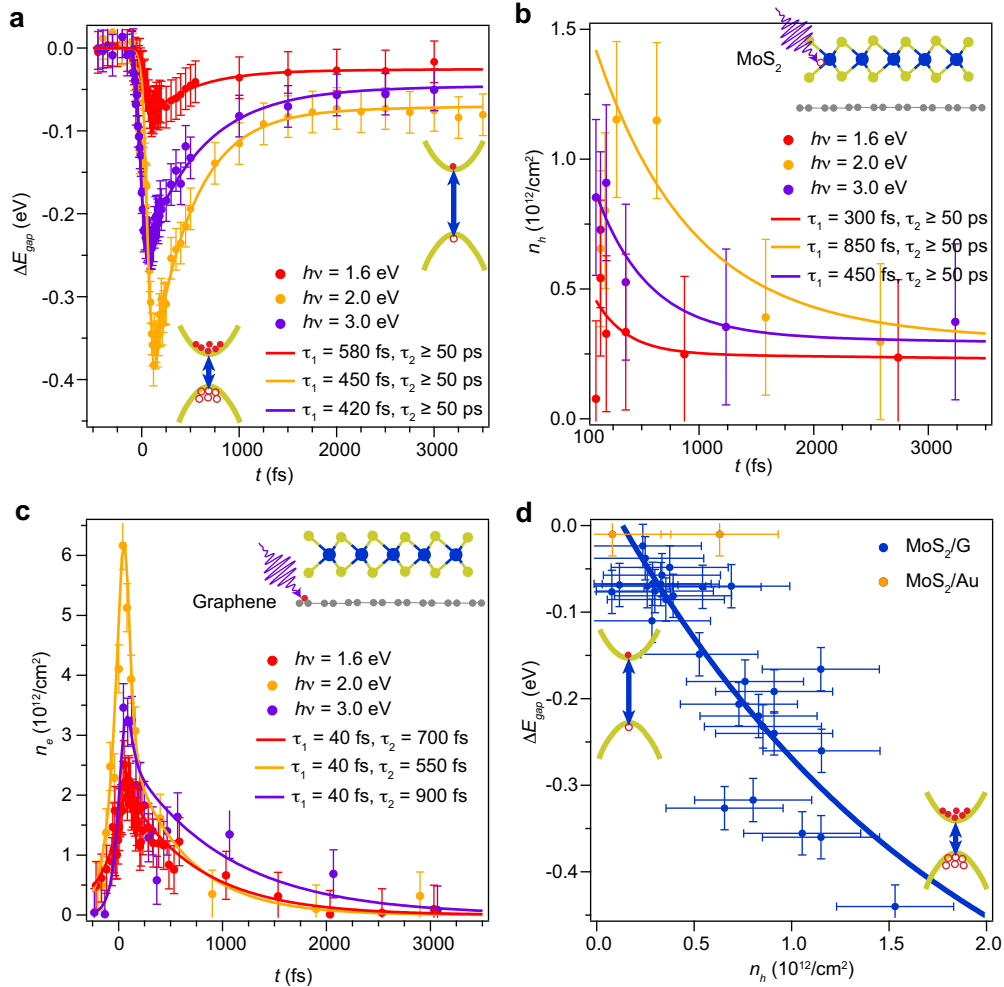}
\caption{Tuning the band gap \textit{via} optically induced carrier densities and substrate screening: (a) Time-dependent band gap renormalization for the data sets in Figure \ref{fig:2}c-e. The lines correspond to an exponential rising edge followed by a double exponential decay with the decay time constants given in the figure. (b) Hole density as a function of delay time for the same data sets. Again, the two time constants for a double exponential decay are given. The time scale here begins at 100~fs because $n_h$ is undefined before the peak optical excitation is reached. For the data after the peak excitation the time-delays have been binned over several time delay points. (c) Time-dependent hole density for the underlying graphene layer, extracted from the data sets in Figure \ref{fig:2}f-h. The time constants for a double exponential decay are given. (d) Plot of the total band gap renormalization \textit{versus} the experimentally extracted hole densities. A comparison with similar data obtained using a Au(111) substrate is included (Data from Ref. \citenum{Antonija-Grubisic-Cabo:2015aa}). The blue curve is merely a guide to the eye.}
\label{fig:5}
\end{center}
\end{figure*}

Figure \ref{fig:2}a shows possible excitation processes in SL MoS$_2$ during the pump phase of the TR-ARPES experiment. For pump pulse photon energies of $h\nu=$1.6~eV and 2.0~eV we observe that the sample is excited even though direct transitions from the VB to the CB are not possible. In these cases, both two-photon absorption or Auger recombination, involving in-gap states, due to defects (illustrated in the left and middle panels of Figure \ref{fig:2}a, respectively) may contribute to excitation. For $h\nu=$2.0~eV where the energy is resonant with the A exciton in MoS$_2$ we expect that exciton dissociation will also become significant leading to excited free electrons and holes.  At $h\nu=$~3.0~eV, direct transitions become possible. For graphene the gapless Dirac cone ensures that direct transitions are possible for all photon energies considered here (Figure \ref{fig:2}b). Excitation of electrons from graphene to MoS$_2$ could be possible in all cases, but we do not observe any indication of this in the time dependence of the signals in the two materials as discussed later in connection with Figure \ref{fig:5}.

Figures \ref{fig:2}c-e and \ref{fig:2}f-h show the intensity difference between a spectrum taken before the optical excitation and a spectrum taken at the peak excitation (typically after 40~fs) for the SL MoS$_2$ and graphene bands. The excitation energy and pump fluence for data sets in the same column are identical, so that the resulting effects can be compared directly. For the MoS$_2$ data we provide calculated values of the optically induced hole density. These calculations are discussed in more depth in connection with Figure \ref{fig:4} and in the Supplementary Material. For graphene, the promotion of electrons from the VB to the CB leads to a depletion of photoemission intensity below the Fermi energy $E_F$  (blue) and a corresponding increase above $E_F$ (red), in the region of the Dirac cone. This behaviour is qualitatively similar to the situation without MoS$_2$ and has been observed for graphene and bilayer graphene for different doping levels \cite{Johannsen:2013aa,Gierz:2013aa,Ulstrup:2014ac,Someya:2014aa,Johannsen:2015}. 

\begin{figure*} [t!]
\begin{center}
\includegraphics[width=0.8\textwidth]{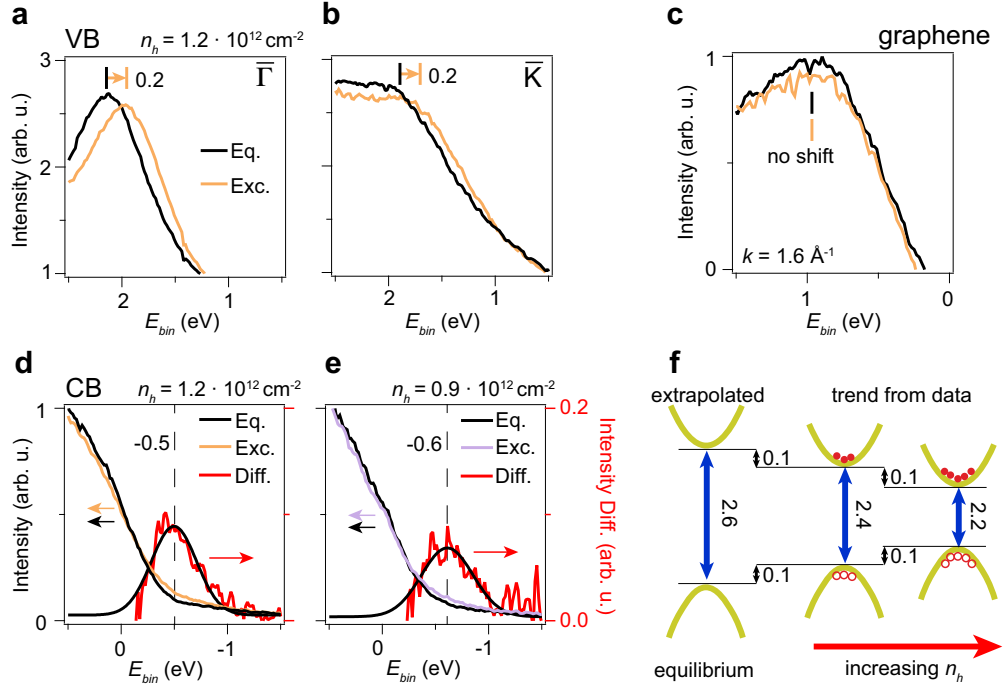}
\caption{Band shifts and band gap renormalization:  (a)-(b) EDC cuts through the data for MoS$_2$ in Figure \ref{fig:2}d at $\bar{\Gamma}$ and $\bar{K}$, respectively, illustrating a rigid shift and an intensity loss in the VB from the equilibrium (Eq.) to the excited (Exc.) state. (c) Corresponding EDC for graphene, taken from the data in Figure \ref{fig:2}g showing just an intensity loss but no shift. (d)-(e) EDCs around $\bar{K}$ in the CB region of MoS$_2$ extracted from the data in Figure \ref{fig:2}d and \ref{fig:2}e, respectively. Curves are displayed in the equilibrium and fully excited cases along with the intensity difference. The peak in the difference is fitted by a Gaussian line shape and the peak value is interpreted as the position of the CBM. (f) Sketch of the observed VB and CB shifts and corresponding band gap renormalization as a function of free carrier density. The shifts are extrapolated to obtain a band gap renormalization of 2.6~eV in the zero free carrier density limit at equilibrium. All numbers are stated in eV unless otherwise noted.}
\label{fig:3}
\end{center}
\end{figure*}

For MoS$_2$ the situation is entirely different: While the CB shows an increased population near the conduction band minimum (CBM), at least for Figure \ref{fig:2}d-e, the VB not only shows the expected intensity loss (blue) but also a pronounced intensity gain (red), immediately above the blue region. This spectral change in the VB corresponds to an overall intensity loss combined with a shift of the band to lower binding energy. We clearly observe that these are rigid band shifts that occur exclusively in the MoS$_2$ related states by analyzing energy distribution curve (EDC) cuts of the data at constant $k$. Figure \ref{fig:3}a and \ref{fig:3}b illustrate the observed rigid shift (0.2~eV) for an optically induced hole density of $1.2\cdot10^{12}$~cm$^{-2}$ at $\bar{\Gamma}$ and $\bar{K}$, respectively. Similar shifts are observed for all excitations but their size depends on the pump energy and fluence, and Figure \ref{fig:2}c-e show data for representative combinations of these parameters. Rigid shifts are shown \textit{via} EDCs at both $\bar{\Gamma}$ and $\bar{K}$ for additional pump pulse settings in Figure S1g-h.

Such spectral shifts can be induced by space charge or surface photovoltage (SPV) effects \cite{Ulstrup:2015j}. The space charge effect is caused by interactions among the photoelectrons in vacuum, which sets up a charge cloud that propagates away from the sample. The acceleration of photoemitted electrons is consequently changed as they propagate to the detector, leading to shifts in the measured kinetic energy distributions. The SPV effect is caused by excitation of electron-hole pairs in the SiC substrate, which modifies the band bending and thereby the electrostatic potential at the surface \cite{Yang:2014aa}. This leads to a time-dependent electrostatic force, exerted by the SPV-derived electric field in the SiC substrate, on all photoemitted electrons, which leads to shifts in the measured spectra. In our case these experimental artefacts can be rigorously excluded since they would affect the graphene bands in the same way as the SL MoS$_2$ bands. This is clearly not the case since shifts for graphene are neither observed in the difference plots of Figure \ref{fig:2}f-h nor in EDC cuts as in Figure \ref{fig:3}c. The timescale of such shifts would resemble the photoelectron propagation time in vacuum, which is several nanoseconds\cite{Tanaka:2012a}. While it is not immediately intuitive, both the SPV and space charge effects cause shifts at negative and positive time delays. At negative time delays, where the photoelectron is emitted before the pump pulse arrives, the electric fields due to pump-induced SPV or space charge clouds can affect the propagating electron in vacuum - an effect which can persist in spectra measured hundreds of picoseconds before the optical excitation as shown for graphene on SiC in Ref. \citenum{Ulstrup:2015j}. At positive time delays, the SPV decays on a timescale corresponding to the carrier recombination time in SiC. In our case, shifts do not occur for negative time delays and the time dependence of the spectral changes at positive time delays does not resemble the photoelectron propagation time in vacuum or the recombination time of excited carriers in SiC (the time dependence of the measured band shifts is shown in Figure \ref{fig:5}).

While the absolute positions and shifts of the VB states may be determined from the EDCs shown in Figure \ref{fig:3}a-b, a corresponding analysis for the CB is limited by the fact that the CB states are not observable in equilibrium, see  Figure \ref{fig:3}d-e. Hence, we can not directly extract the initial binding energy of the CBM. We do, however, observe indications of excited CB states above the Fermi level as an intensity increase around $\bar{K}$ following photoexcitation, at least for a sufficiently strong excitation as is the case in  Figures \ref{fig:3}d-e. For these datasets we can fit a Gaussian line shape to an EDC at $\bar{K}$, and use the fitted peak position as a CBM estimate. In Figure \ref{fig:2}d-e a slight intensity loss is visible immediately below the Fermi level. The spectral weight in this region comes from in-gap defect states. The optical excitation produces a distribution of hot electrons (and holes) in these states, as illustrated in all panels of Figure \ref{fig:2}a. The broadening of the Fermi edge that accompanies this hot carrier population is directly visible in Figure \ref{fig:3}d-e. The surplus of intensity above the Fermi level due to these defect states contributes as a background to the main signal coming from the excited population in the CBM. From comparing the CBM position as a function of photon energy and fluence in Figures \ref{fig:3}d-e, we find that the CBM shift is the same size but opposite direction as the observed shifts of the VBM in Figures \ref{fig:3}a-b and S1g-j.  Even though the uncertainties are large in this analysis, extrapolating such symmetric VBM and CBM shifts to zero excited carrier density leads to an equilibrium quasiparticle gap consistent with a value of 2.6~eV as illustrated in Figure \ref{fig:3}f. Note that the overlaid calculated CB and VB in Figure \ref{fig:2}c-e have been rigidly shifted in accordance with this analysis.

\begin{figure} [t!]
\begin{center}
\includegraphics[width=1\textwidth]{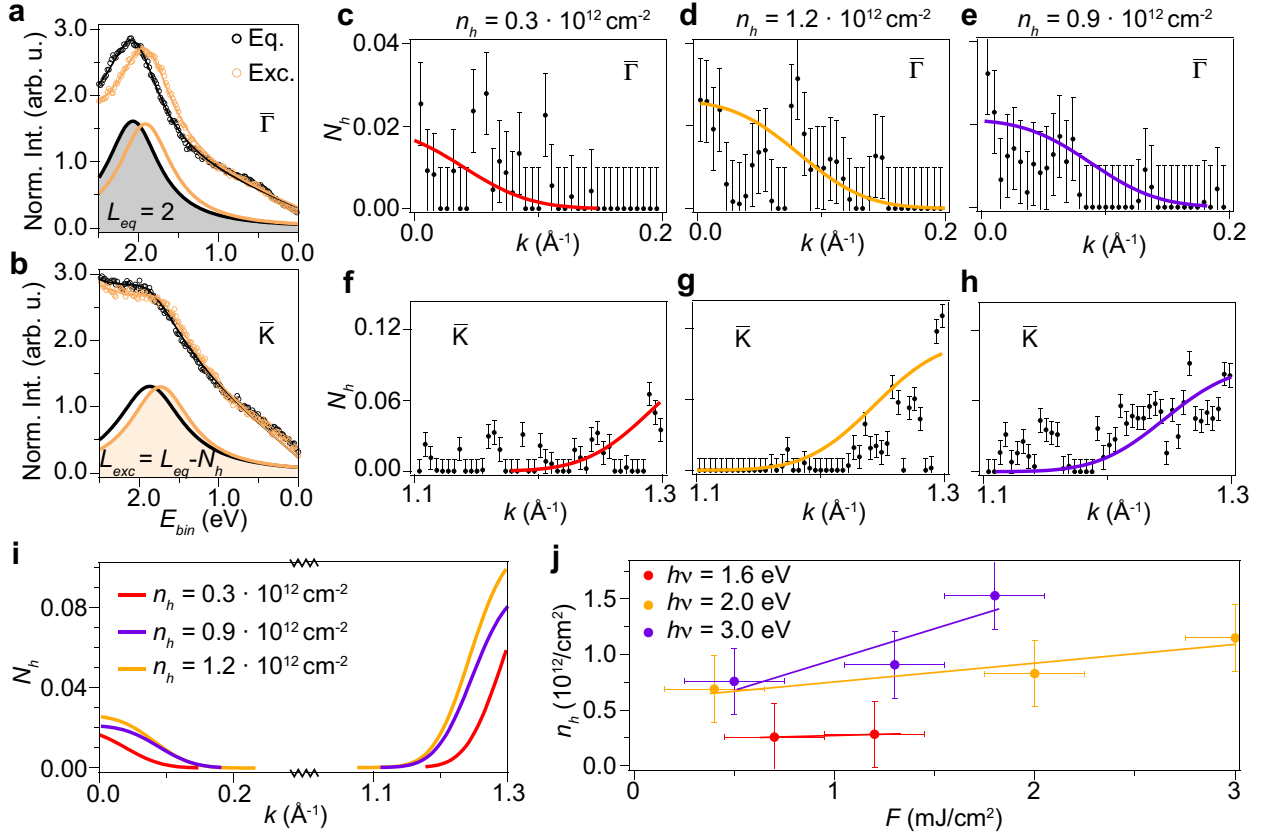}
\caption{Analysis of optically induced hole density in the SL MoS$_2$ layer: (a)-(b) EDCs (markers) at $\bar{\Gamma}$ and $\bar{K}$, plotted for a time delay before the optical excitation (black) and at peak excitation (orange) for the data in Figure \ref{fig:2}d. Fits to a linear background and a single Lorentzian line shape are shown by lines and the Lorentzian parts are plotted individually. The data are normalized such that the Lorentzian area $L_{eq}$ before excitation integrates up to 2 electrons ($L_{eq}=2$, shown for $\bar{\Gamma}$ in (a)). The number of generated holes in the state $N_h$ can then be determined from the Lorentzian area in the excited case  $L_{exc}$ by $N_h=L_{eq}-L_{exc}$ (shown for $\bar{K}$ in (b)). (c)-(e) Hole profiles $N_h(k)$ around $\bar{\Gamma}$ and (f)-(h) around $\bar{K}$. The markers are the experimental data obtained from the analysis in (a)-(b). Smooth curves are fits to a distribution function. (i) Fitted distribution functions from (c)-(h) shown for clarity. (j) Calculated hole density for SL MoS$_2$ at peak excitation for different combinations of pump energy and fluence. Lines through the data points have been added as a guide to the eye.}
\label{fig:4}
\end{center}
\end{figure}

In order to determine the optically induced number of holes that accompanies the band shifts, we apply an EDC-based analysis of the intensity loss in the VB introduced in Figure \ref{fig:4}(a)-(b). As shown for EDCs around the $\bar{\Gamma}$ (Figure \ref{fig:4}a) and $\bar{K}$ (Figure \ref{fig:4}b) points for the data in Figure \ref{fig:2}d, we perform a fit with a Lorentzian on a linear background. The background is allowed to vary for each momentum point due to tails from nearby graphene bands, however, the background is held fixed as a function of time. The data taken before the optical excitation are normalized such that the Lorentzian integrates up to the number of electrons (2), which occupy a single momentum state in equilibrium. The integrated Lorentzian after the optical excitation can then be used to determine the number of holes in the $k$-range of the EDC. Applying this procedure for the data along $\bar{\Gamma}$-$\bar{K}$ results in the number of holes $N_h$ at peak excitation as a function of $k$, as shown in Figure \ref{fig:4}c-h for the same pump pulse settings as the corresponding Figure \ref{fig:2}c-e. The overall intensity loss in the EDCs following excitation is comparable to the level of the noise, however, in the data $N_h$ is strongly peaked around the VB maxima at $\bar{\Gamma}$ and $\bar{K}$  at any given time and clearly approaches zero away from the band maxima at $\bar{\Gamma}$ and $\bar{K}$. This behavior agrees with the expectation, that the holes reach a quasi-equilibrium on a much faster time scale than our time resolution of 40~fs, leading to transient Fermi-Dirac like distributions near the band extrema. Fits of such functions (see Figure S3c for an example of a raw fitting function) to all data are shown in each panel of Figure \ref{fig:4}c-h and combined in Figure \ref{fig:4}i for ease of comparison. The number of holes never exceeds 10\% of the available states.

Because of the hole distribution's strong energy dependence and the tendency to even an anisotropic distribution out on a short time scale \cite{Mittendorff:2014aa}, it is reasonable to assume an isotropic distribution around the local band maxima.
The total hole density $n_h$ in the VB can then be calculated and the result is shown at peak excitation in Figure \ref{fig:4}j. As expected, $n_h$ depends on both the pump photon energy and fluence. The lines added to each set of points in Figure \ref{fig:4}j have been included as a guide to the eye to compare the overall magnitude and trend of hole density with fluence for a given pump pulse energy. The excited hole density is expected to extrapolate to zero and thus be a highly nonlinear function with fluence. A complete understanding of this dependence is beyond the scope of this work. Furthermore, a simple estimate using the pump pulse fluences and photon energies applied here would lead to excited hole densities on the order of $10^{14}-10^{15}$~cm$^{-2}$ using the absorbance of 5-10~\% for SL MoS$_2$ \cite{bernardiextraordinary2013}. The much smaller hole densities we obtain are consistent with previous TR-ARPES studies on graphene\cite{Johannsen:2013aa,Johannsen:2015}, and can be attributed to a combination of effects such as Pauli blocking, recombination of carriers on faster timescales than we probe, diffusion of excited carriers out of the probed area of the sample, as well as imperfect overlap of pump and probe beams. Note that  Figures \ref{fig:2} and \ref{fig:4} only provide values at peak excitation but the energy shifts and hole densities are analyzed for entire sets of data with different delay times between pump and probe pulse as shown in Figure S1i-j. This allows us to correlate hole densities with the corresponding band gap reduction for a wide range of parameters and conditions.

In Figure \ref{fig:5}a we analyze the dynamical band gap reduction as a function of delay time between the pump and probe pulse for the three data sets from Figure \ref{fig:2}c-e. The largest effect of a nearly 0.4~eV shrinkage is seen for the data taken with a pump pulse energy of 2~eV, which also has the highest fluence and results in the highest hole density at peak excitation (see Figure \ref{fig:2}d). The decay of the band gap shrinkage is fitted with a double exponential function and the characteristic decay times are similar for the three data sets ($\approx 500$~fs and $\approx 50$~ps). Note that the time constant of 50~ps for the slow component is a lower estimate. Figure \ref{fig:5}b shows the corresponding time dependence of the hole density. A double exponential fit to these results in similar time constants but the uncertainties on the individual data points are very large, even when wider ranges of time delays are binned into a single data point. Nevertheless, the similar time constants strongly support the interpretation that the observed band narrowing is induced by the excited carriers. The presence of two characteristic decay times is also consistent with all-optical experiments \cite{wangultrafast2015}.

Figure \ref{fig:5}c shows the time-dependent electron density in the graphene layer for the three data sets shown in Figure \ref{fig:2}f-h, illustrating that the carrier dynamics is entirely different from the dynamics of the band gap reduction and hole density in the SL MoS$_2$. Here a double exponential fit results in a fast decay time that is more than an order of magnitude shorter than that observed for SL MoS$_2$. Since the optically induced electrons in graphene thermalize within our time resolution, it is possible to estimate the transient electron density from the electronic temperature. The extracted time-dependence of the electron density corresponds well to that observed in previous experiments without the SL MoS$_2$ on top\cite{Someya:2014aa,Johannsen:2015}. Even though we have regions of bare graphene within the probed area of the sample due to our MoS$_2$ coverage of approximately 0.55~ML\cite{miwavander2015} we also expect some signal from the graphene under the MoS$_2$\cite{zhangdirect2014}. The observation of a single time constant suggests that the graphene dynamics are at most weakly affected by the overlaid MoS$_2$.

Finally, Figure \ref{fig:5}d summarizes the key result of this paper, the band gap shrinkage as a function of excited hole density, combining data taken with different excitation energies and fluences. These results can be compared to a recent prediction by Steinhoff \emph{et al.} which gives a somewhat larger shrinkage but with the same order of magnitude as observed here ($\approx 0.5$~eV for a carrier density of $\approx 10^{12}$~cm$^{-2}$) \cite{Steinhoff:2014aa}. The discrepancy between experiment and theory can easily be accounted for by the uncertainties in the determination of the carrier concentration and the role of the underlying graphene, which is not present in the calculation.

Similar carrier-induced renormalization effects have been observed for quasi-free-standing MoS$_2$ on SiO$_2$, either by optical or electrical doping \cite{Chernikov:2015aa,Chernikov:2015ab} and for surface doping of bulk WSe$_2$ \cite{Riley:2015aa}. However, the effect is nearly completely suppressed when MoS$_2$ is placed in a strongly screening environment such as on a metallic Au(111) substrate \cite{Antonija-Grubisic-Cabo:2015aa,Bruix:2016}. To illustrate this, we have added  the result of a corresponding analysis, using data for MoS$_2$/Au(111) from Ref. \citenum{Antonija-Grubisic-Cabo:2015aa}. In this case (orange data points in Figure \ref{fig:5}d) the band shift remains below 10~meV, even for high induced carrier densities. Also, the typical decay times for excited carriers ($\approx$50~fs) are much faster than observed here or for MoS$_2$ on SiO$_2$ \cite{wangultrafast2015,Chernikov:2015aa}. Our results are reminiscent of TR-ARPES studies on the strongly correlated metallic TMDC 1T-TiSe$_2$\cite{Rohwer:2011,Hellmann:2012}. This system exhibits a charge ordered state, which is completely removed due to additional screening from optically excited free carriers\cite{Rohwer:2011}. This scenario is consistent with our interpretation of a band gap reduction in SL MoS$_2$ due to the build-up of screening \textit{via} laser-induced free carriers.

\section{Conclusion}

As our results show, the static screening of charge carriers in SL MoS$_2$ in a heterostructure with graphene is sufficiently weak that a significant band gap renormalization can be induced by optically excited carriers. On a fundamental level, this suggests that the degree of control of the electronic structure of such heterostructures extends beyond the substrate-induced renormalization of the static band gap of the 2D semiconductor to the dynamic tuneability of the band gap and the electron dynamics. This behavior is essential to take into account in optical applications where the change of the resonance condition on ultrafast time scales may pose a serious challenge, \textit{e.g.} in the design of an actual optical cavity. On the other hand, one could imagine optoelectronic devices based on a heterostructure with different band alignments in the constituent materials where ultrafast light-induced switching of the band offsets could be used as a means to control the separation of excitons and free charge carriers.

\section{Materials and Methods}

\subsection{Sample preparation}
The MoS$_2$/G heterostructure was synthesized on a 6H-SiC 2 in. wafer (TanKeBlue Semiconductor Co. Ltd., $n$-type doping, 0.02-0.10~$\Omega\cdot$cm). The epitaxial graphene layer was grown first on the Si terminated (0001) face by direct current annealing of the  6H-SiC substrate in a dedicated ultra-high vacuum (UHV) graphene growth chamber with a base pressure of $5\times10^{-10}$~mbar. The dangling Si bonds of the substrate were saturated by a carbon buffer layer with a $(6\sqrt{3} \times 6\sqrt{3})R30^{\circ}$ periodicity under the graphene layer. MoS$_2$ was then grown on top of the graphene by van der Waals epitaxy in a separate UHV chamber. In this procedure the Mo was initially deposited from an electron beam evaporator (Oxford Instruments) in a H$_2$S atmosphere of $\approx1\times10^{-5}$~mbar. After deposition the sample was annealed for $\approx1$~hour at 1050~K in the H$_2$S environment, which resulted in nanosized islands of MoS$_2$ on the graphene. Repeated cycles of this process lead to the formation of a SL MoS$_2$ film with an optimum coverage around 0.55~ML before bilayer MoS$_2$ islands nucleated, as determined by scanning tunneling microscopy. Further details on the growth are provided in Ref. \citenum{miwavander2015}.\\ 

\subsection{ARPES}
The equilibrium electronic structure of the MoS$_2$/G sample was determined by ARPES at the SGM-3 UHV end-station of the synchrotron radiation source ASTRID2 in Aarhus, Denmark \cite{Hoffmann:2004aa}. The sample was initially annealed to 500~K to remove  adsorbed species. The ARPES data was collected at a photon energy of 70~eV over a significant portion of the MoS$_2$ and graphene Brillouin zones (BZs) in order to determine the location of the band extrema of the two materials and their relative orientation with respect to each other. This information was used to align the sample in the TR-ARPES experiments. The sample temperature was kept at 70~K and the total energy- and angular-resolution were set to 20~meV and 0.2$^{\circ}$, respectively.\\

\subsection{TR-ARPES}
The same sample that was measured by ARPES was then transported in an evacuated tube pumped down below $1~\times~10^{-9}$~mbar to the Artemis facility, Rutherford Appleton Laboratory in Harwell, UK for TR-ARPES measurements \cite{Frassetto:2011}. Here it was placed in the TR-ARPES UHV end-station and annealed to 500~K in order to remove any adsorbed surface contaminants. The sample temperature was kept at 50~K using a liquid helium cryostat during measurements.

The pump and probe pulses used for acquiring TR-ARPES spectra were generated using a 1~kHz Ti:sapphire amplified laser system with a wavelength of 785~nm, a pulse duration of 30~fs and an energy per pulse of 12~mJ. The band structures of MoS$_2$ and graphene were measured along their $\bar{\Gamma}$-$\bar{K}$ directions using high harmonic probe pulses of $h\nu=25$~eV, which were generated by focusing a part of the laser energy on a pulsed jet of argon gas. The remaining laser energy was applied to drive an optical parametric amplifier (HE-Topas) followed by a frequency mixing stage. This provided tunable pump pulses with wavelengths centered at 408~nm (3.0~eV), 615~nm (2.0~eV) and 784~nm (1.6~eV), which were used to optically excite the sample. The fluence of the pump pulse was kept in the range 0.4-3.0~mJ/cm$^{2}$, and the maximum used for any of the three applied wavelengths was kept below the onset where significant energy broadening and shifts of the measured spectra started to occur due to pump-induced space-charge effects \cite{Ulstrup:2015j}. The beams were polarized such that the pump pulse was $s$-polarized and the probe pulse was $p$-polarized. 

The time delays of pump and probe pulses were varied using a mechanical delay line. We applied two modes of acquisition: (1) A few-delay-point mode where we measured the spectra over 3-6 time delay points around peak optical excitation but acquired these cumulatively up to 2500 times in order to optimize signal-to-noise ratio. (2) A many-delay-point mode, where we measured the spectra over 50-60 time delay points cumulatively up to 1000 times. Acquisition times of the few-delay-point mode were up to 3~h, which enabled us to acquire data at 3-4 different pump fluence settings per pump wavelength. In the many-delay-point mode acquisition times were up to 20~h per dataset, which limited the number of experimental parameters that we could vary in such experiments, but allowed us to capture the detailed time dependence. Throughout this work we have defined $t=0$ to coincide with the middle of the rising edge in the measured spectra following optical excitation of the sample. The energy-, angular- and time-resolution were 400~meV, 0.3$^{\circ}$ and 40~fs, respectively. 

\begin{acknowledgement}
We thank Phil Rice for technical support during the Artemis beamtime. We gratefully acknowledge funding from the VILLUM foundation, the Lundbeck foundation, EPSRC (Grant Nos. EP/I031014/1 and EP/L505079/1), The Royal Society and the Swiss National Science Foundation (NSF). Ph. H. and S. U. acknowledge financial support from the Danish Council for Independent Research, Natural Sciences under the Sapere Aude program (Grant Nos. DFF-4002-00029 and DFF-4090-00125). Access to the Artemis Facility was funded by STFC. Data underpinning this publication can be accessed at \url{http://dx.doi.org/10.17630/2898c6e7-00ba-44fe-9ca2-cb08501a6bca}.
\end{acknowledgement}

\providecommand{\latin}[1]{#1}
\providecommand*\mcitethebibliography{\thebibliography}
\csname @ifundefined\endcsname{endmcitethebibliography}
  {\let\endmcitethebibliography\endthebibliography}{}

\end{document}